\documentclass[reprint, twocolumn,superscriptaddress,longbibliography]{revtex4-2}
\usepackage[T1]{fontenc}
\usepackage[version=4]{mhchem}
\usepackage{graphicx}
\usepackage{placeins}
\usepackage{verbatim}
\usepackage[usenames, dvipsnames]{xcolor}
\usepackage{algpseudocode}
\usepackage{hyperref}
\date{\today}
\begin{document}
\title{Towards Structural Reconstruction from X-Ray Spectra}
\author{Anton Vladyka}
\email{anton.vladyka@utu.fi}
\affiliation{University of Turku, Department of Physics and Astronomy, 20014 Turun yliopisto, Finland}
\author{Christoph J.~Sahle}
\email{christoph.sahle@esrf.fr}
\affiliation{European Synchrotron Radiation Source, 71 Avenue des Martyrs, 38000 Grenoble, France}
\author{Johannes Niskanen}
\email{johannes.niskanen@utu.fi}
\affiliation{University of Turku, Department of Physics and Astronomy, 20014 Turun yliopisto, Finland}

\begin{abstract}
We report a statistical analysis of Ge K-edge X-ray emission spectra simulated for amorphous GeO$_2$ at elevated pressures. We find that employing machine learning approaches we  can reliably predict the statistical moments of the K$\beta''$ and K$\beta_2$ peaks in the spectrum from the Coulomb matrix descriptor with a training set of $\sim 10^4$ samples. Spectral-significance-guided dimensionality reduction techniques allow us to construct an approximate inverse mapping from spectral moments to pseudo-Coulomb matrices. When applying this to the moments of the ensemble-mean spectrum, we obtain distances from the active site that match closely to those of the ensemble mean and which moreover reproduce the pressure-induced coordination change in amorphous GeO$_2$. With this approach utilizing emulator-based component analysis, we are able to filter out the artificially complete structural information available from simulated snapshots, and quantitatively analyse structural changes that can be inferred from the changes in the K$\beta$ emission spectrum alone.
\end{abstract} 
\maketitle

\section{Introduction}
Core-level spectroscopy provides information of structure of matter at the atomic level, and the constituent methods are applied from standard material characterization to conceptually new experiments at large-scale facilities such as free-electron lasers. Although reference data helps, interpretation of core-level spectra is not always straightforward, especially in the case of soft condensed or amorphous matter where ensemble statistics plays a drastic role \cite{Wernet2004,Ottosson2011,Niskanen2016a,niskanen2017disentangling,niskanen2019compatibility,VazDaCruz2019,Pietzsch2022}. Studies of this statistical nature, and the implied repeated function evaluations, could benefit from machine learning (ML), application of which to core-level spectra has been studied rather intensively lately \cite{Timoshenko2018, Timoshenko2019,Carbone2019,Carbone2020, Rankine2020, Guda2021,Niskanen2022neural,Niskanen2022}. In general, when working with atomic resolution studies have raised the need to engineer features for both structure \cite{Rupp2012,Bartok2013,Timoshenko2017,Ghosh2019,Langer2022} and spectra \cite{Ghosh2019,Guda2021}.
\par
The pressure dependent evolution of the germanium coordination by oxygen in glassy GeO$_2$ has been a long standing subject of study \cite{guthrie2004formation, lelong2012evidence, kono2016ultrahigh,Spiekermann2019}. Besides applications of amorphous GeO$_2$ in technical glasses, the increased sensitivity of a-GeO$_2$ to pressure compared to amorphous SiO$_2$ motivates the study of structural changes similar to those expected to occur in the pressurized analogue glass a-SiO$_2$ but at greatly reduced absolute pressures. Detailed knowledge of the compaction mechanisms in these simple glasses will have direct consequences for our understanding of geological, geochemical, and geophysical processes involving more complex silicate glasses and melts.
\par
X-ray emission spectra (XES) of GeO$_2$ is an inviting case for development of spectroscopic analysis for soft and amorphous condensed matter. First, large spectroscopic changes with changing local structure are known to exist \cite{Spiekermann2019}. Second, simulations are known to reproduce the observed ensemble-mean effects well\cite{Spiekermann2022}. Third, XES is local-occupied-orbital derived and a few orbital-bonding neighbor atoms are expected to be decisive for the spectrum outcome. This would result in a minimal set of structural parameters needed to predict XES. Last, owing to the chemical simplicity and simple bonding topology due to non-molecular structure, this system has promise to be reproduced by ML with the limited number of data points that the condensed phase allows. Namely, for such systems the electronic simulation needs to account for multi-electron effects in numerous interacting atoms -- typically on the level of density functional theory. As a consequence, the number of individual structural data points for spectroscopy can be expected to be $\sim$10$^4$ in an extensive contemporary simulation.
\par
In this work, we focus on Ge K$\beta$ XES calculations of amorphous GeO$_2$ at elevated pressures. Our previous work on the water molecule indicated that predicting spectral features is easier than predicting structural features \cite{Niskanen2022neural}. In the condensed phase, where the structural features to be predicted are more numerous, the task is arguably even more complicated. As a solution to this dilemma, we build a procedure on spectrum prediction for structures, dimensionality reduction and iterative optimization algorithms. This approach is possible because the evaluation of an ML model requires much less computational resources than the corresponding quantum mechanical calculation does. We predict statistical moments of XES lines from a Coulomb matrix\cite{Rupp2012} that describes the local atomic structure around the site of characteristic X-ray emission. Next, we study obtainable structural information for the occurring spectral changes in the pressure progression of the XES by emulator-based component analysis (ECA)\cite{Niskanen2022}. Last, we investigate an approximate solution to the spectrum-to-structure inverse problem by first transforming it into an optimization task in the dimension-reduced ECA space, followed by expansion to the full multi-dimensional Coulomb matrix. A dedicated evaluation data set allows for assessment of performance of the approach in each of the aforementioned tasks.
\section{Methods}
After ionisation from a Ge 1s orbital, the electronic system is left in a highly excited state, which decays by either Auger decay or by emission of a photon, accompanied by a transition of an electron from a higher-energy orbital. For germanium, transitions from 3p to 1s orbital give rise to so called K$\beta$ emission spectra. Since the Ge 3p orbitals constitute valence orbitals, Ge K$\beta$ XES is highly sensitive to chemical bonding of the active Ge site.

We study data of amorphous GeO$_2$ from statistical spectral simulations over a range of 11 pressure values from 0 GPa to 120 GPa. These XES spectra, simulated using the OCEAN code \cite{vinson2011bethe, gilmore2015efficient} (version 2.5.2), are based on real-space configurations from \textit{ab initio} molecular dynamics simulations reported earlier by Du \textit{et al.}~\cite{du2017oxygen}.
We used the Quantum ESPRESSO program package (version 5.0) \cite{giannozzi2009quantum, QEwebsite} for sampling the ground state wave functions and electron densities at the gamma point with a plane wave cutoff of 100 Ry (see Ref.~\citenum{Spiekermann2022} for more details). Transition matrix elements are then calculated using the Haydock recursion method \cite{haydock1975electronic} as implemented in the OCEAN code using a Lorentzian width of 1.0 eV for the continued fraction. At each pressure point, Ge K$\beta$ XES spectra of 18 structurally uncorrelated AIMD simulation snapshots containing 72 GeO$_2$ formula units were calculated for each Ge atom. For 5 pressure points only 17 out of 18 snapshots yielded spectra in a finite time frame due to convergence issues, resulting in 13896 XES spectra. The spectra of individual Ge sites were aligned and normalized for each pressure to yield a constant K$\beta_5$ line peak position and intensity in its ensemble average spectrum.
\par
Even though extensive from a statistical simulation viewpoint, the available dataset is still rather limited for sophisticated ML algorithms. In this case, using a descriptive numerics allows for condensing structural and spectral information to a few parameters, resulting in an improvement of ML performance. We apply descriptors to both the spectrum and the atomic structure of the system (see below). 

\subsection{Spectral-line descriptor}
The XES spectrum is given as a function presenting intensity against photon energy in eV ($I = I(E)$). For a distinguishable peak in the spectrum, we use raw moments defined as follows:
\begin{eqnarray}
M_1 &=& \frac{\int I(E)E\mathrm{d}E}{\int I(E)\mathrm{d}E} ,\\
M_n &=& \frac{\int I(E)(E-M_1)^n\mathrm{d}E}{\int I(E)\mathrm{d}E},\qquad\textrm{for } n = 2,3,4.
\end{eqnarray}
Corresponding spectral descriptors used in the model are spectrum peak position mean $\mu = M_1$ (eV), standard deviation $\sigma = \sqrt{M_2}$ (eV), skewness $\mathrm{sk} = M_3/\sigma^3$ and excess kurtosis $\mathrm{ex} = M_4/\sigma^4 - 3$.  These descriptors are referred to as ``spectral moments'', and are presented as a vector $\mathbf{m}$ throughout the manuscript. In this work, 4 moments were calculated for both K$\beta''$ and K$\beta_2$ peaks, which resulted in 8 descriptors per spectrum.

\subsection{Structural descriptors}
\begin{figure}[t]
    \centering
    \includegraphics[width=\linewidth]{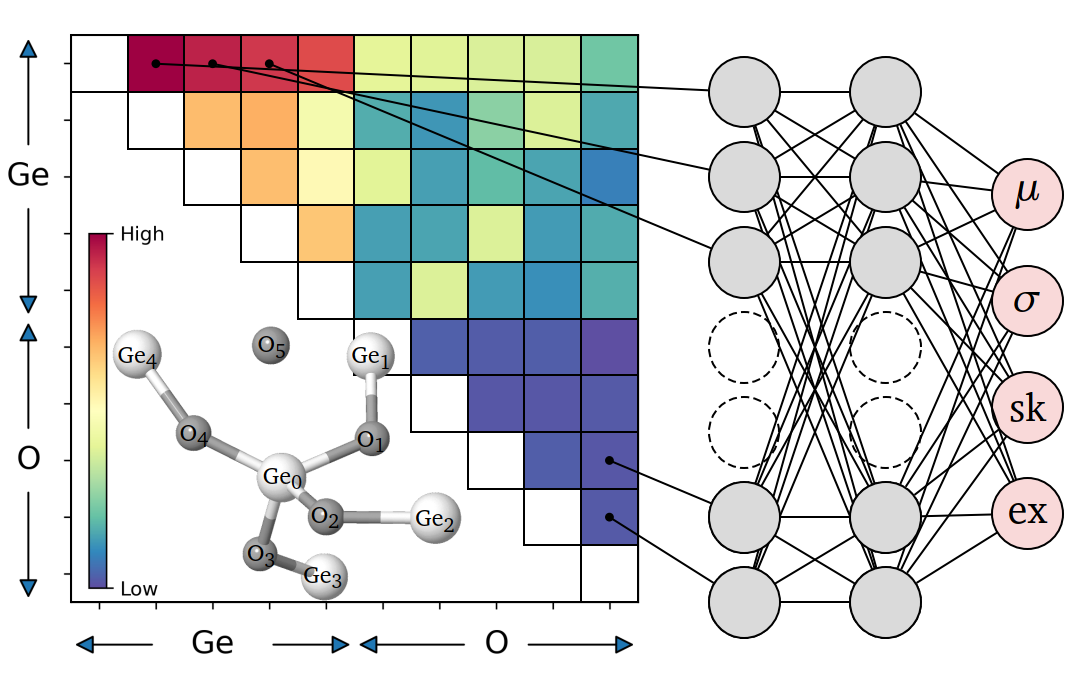}
    \caption{The principle of spectral moment prediction for a Ge K$\beta$ XES peak for amorphous GeO$_2$. A Coulomb matrix is generated from a structure, and its upper triangle is fed as input for MLP, which is trained to predict spectral moments of the line of interest.}
    \label{fig:cmatrix}
\end{figure}

As the structural descriptor we use the Coulomb matrix \cite{Rupp2012}, the elements of which are defined as
\begin{equation}
C_{ij} = \begin{cases} 0.5Z_i^{2.4}\quad\text{ if } i = j, \\ \frac{Z_iZ_j}{R_{ij}}\quad\text{ if } i \ne j,\\
            \end{cases}
\end{equation}
where $Z_i$ is the atomic number of the $i$-th atom, and $R_{ij}$ is the distance between the $i$-th and $j$-th atoms. In this work, we arrange the atoms by the distance from the active site in ascending order, and grouped Ge atoms first, followed by the O atoms. Since in this approach the order of atoms is the same for a given number of Ge and O atoms, the diagonal elements of the Coulomb matrix are identical for each structure. Therefore, owing to symmetry of the matrix, only the upper triangle of the Coulomb matrix is used (see Fig.~\ref{fig:cmatrix}) as vector $\mathbf{p}$ as an input data. From an optimization search, we deduced the optimal number of the atoms used for the Coulomb matrix calculation to be the 10 closest Ge and 7 closest O atoms, which leads to 153-dimensional feature vectors (see details in SI, Fig.~\ref{fig:features}). 
\par
The definition of the Coulomb matrix implies that it can be inverted to a distance matrix containing interatomic distances by
\begin{equation}
R_{ij} = \frac{Z_iZ_j}{C_{ij}}\quad\text{ with } i \ne j,
\end{equation}
where $R_{ij}$ is the distance between atoms $i$ and $j$. This conversion is possible as the $Z_i$ of the chemical elements in each matrix element $C_{ij}$ is known. Furthermore, apart from the handedness of the coordinate system, the atomic geometry can be reconstructed from the distance matrix $\mathbf{R}$, and therefore, from the Coulomb matrix $\mathbf{C}$ (see Supplementary Information for algorithm).
\par
To check the performance of the Coulomb matrix descriptor against a many-body-tensor-representation\cite{Huo2022} spirited descriptor, we used snapshot-wise evaluated radial distribution functions (RDF) from the active site. 
Although similar predictive power was obtained via the RDF, its performance in the later steps of the analysis (spectral coverage of decomposition) was inferior to that of the Coulomb matrix.

\subsection{Algorithms}
The structural and spectral data are presented as feature-wise standardized matrices $\tilde{\mathbf{P}}$ and $\tilde{\mathbf{M}}$, respectively (individual data points $\tilde{\mathbf{p}}$ and $\tilde{\mathbf{m}}$ occupy rows in these matrices). The analysis algorithms aim at discovering the correlations between the two data sets, for which we first applied the emulator-based component analysis (ECA) method as described in Ref.~\citenum{Niskanen2022}. This algorithm relies on a machine-learning based emulator for spectral features at a vector of structural descriptors, that may be previously unseen to it. The algorithm uses projection of structural data on a subspace so that projected data maximize the generalized covered spectral variance (R$^2$ score) when a prediction is made using the emulator. Here, ECA is applied to standardized Coulomb matrix parameters $\tilde{\mathbf{p}}$ and the corresponding standardized spectral moments $\tilde{\mathbf{m}}$. The decomposition algorithm results in orthonormal standardized-structural-parameter-space vectors $\tilde{\mathbf{v}}_1, \tilde{\mathbf{v}}_2,\ldots$ so that spectral moments $\tilde{\mathbf{m}}_\mathrm{emu} = \mathbf{S}_\mathrm{emu}(\tilde{\mathbf{p}}^{(k)})$ for projections
\begin{equation}\label{eq_projection}
\tilde{\mathbf{p}}^{(k)}=\sum_{i=1}^{k} \tilde{\mathbf{v}}_i \underbrace{(\tilde{\mathbf{v}}_i\cdot \tilde{\mathbf{p}})}_{=:t_i},
\end{equation}
predicted using trained neural network $\mathbf{S}_\mathrm{emu}$, cover most of spectral variance of the respective set of points $\tilde{\mathbf{p}}$ at the given rank $k$. Scores $t_i$ are coordinates of the approximate point $\tilde{\mathbf{p}}^{(k)}$ in the $k$-dimensional subspace.

The ECA method requires an emulator capable of predicting spectral moments for new structural data points. As an emulator, a trained multilayer perceptron (MLP) with 2 hidden layers and 64 neurons in each layer was used. We dedicated 80\% of data for training, and 20\% for evaluation of the prediction. Overall, all configurations of MLPs with 2--3 hidden layers and 64 or 128 neurons were evaluated on the training dataset ($\sim$11000 spectra) using mean squared error as a training metric.
\par
For comparison, we used partial least squares fitting based on singular value decomposition (PLSSVD) \cite{booksteinPLSSVD1996} as applied to the X-ray spectroscopic problem in Ref.~\citenum{Niskanen2022}. The PLSSVD algorithm relies on projections of spectral and structural feature vectors on latent variables between which a linear fit is made. The method results in an approximation of the data up to rank $k$
\begin{equation}
	\label{eq:plssvd}
	\tilde{\mathbf{M}} \approx \tilde{\mathbf{P}}\sum_{i=1}^{k} U^{(i)} c_i V^{(i)\mathrm{T}},
\end{equation}
where $U^{(i)}$ is the $i$-th (column) basis vector of the structural descriptors and $V^{(i)}$ is the $i$-th (column) basis vector of the spectral descriptor space. The coefficient $c_i$ is obtained by a fit to the scores $(\tilde{\mathbf{P}}U^{(i)}, \tilde{\mathbf{M}}V^{(i)})$. The orthonormal basis vectors are obtained from a singular value decomposition of the covariance matrix of the data $\mathrm{cov}(\tilde{\mathbf{P}},\tilde{\mathbf{M}})=\tilde{\mathbf{P}}^\mathrm{T}\tilde{\mathbf{M}}$; ordering along descending magnitude of the singular values $\lambda_i$ is applied. For analysis using the PLSSVD algorithm, the same evaluation data set  as for ECA was used.

\begin{figure}
    \centering
    \includegraphics[width=\linewidth]{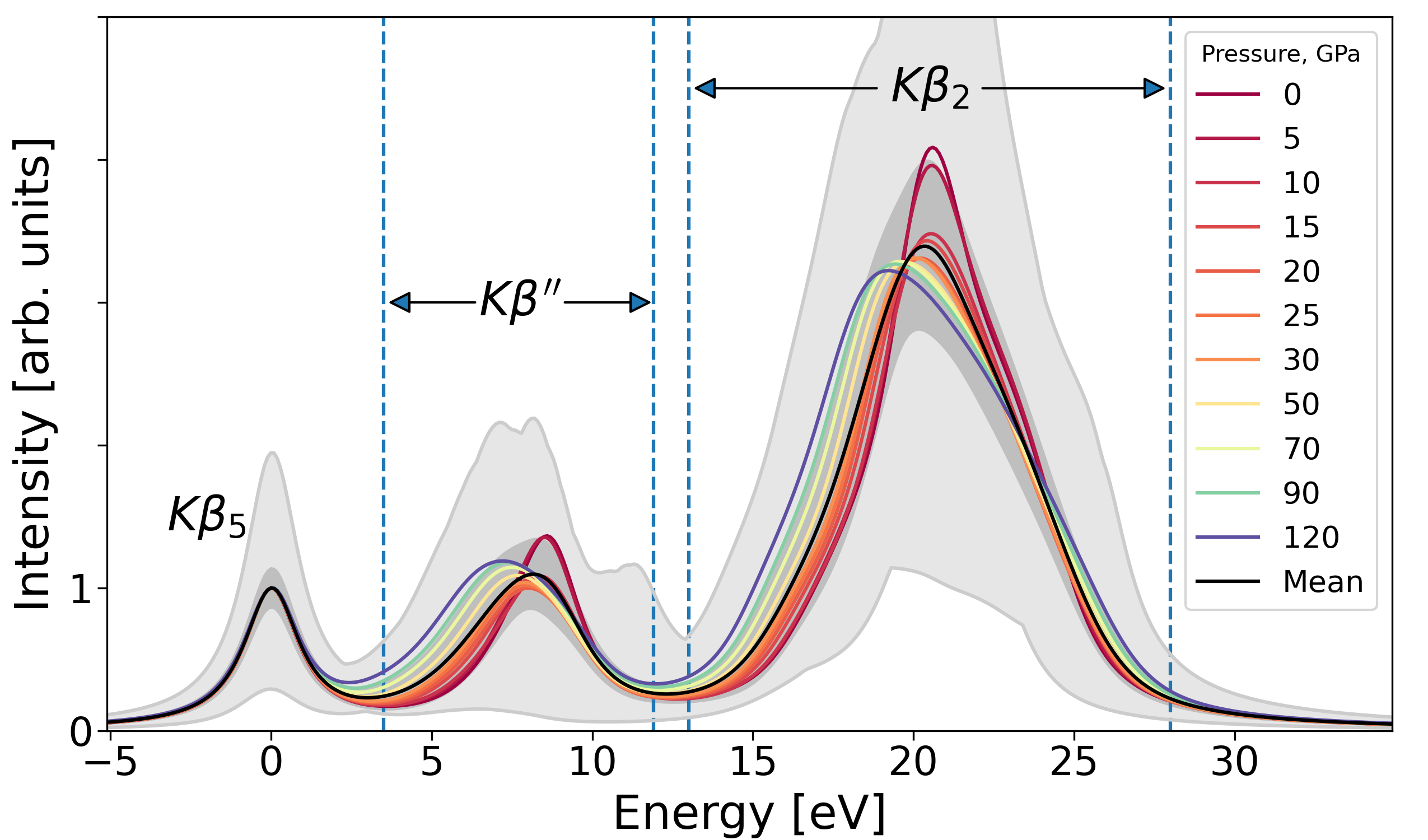}
    \caption{Raw XES spectra. Colored curves depict the mean spectra for each pressure, black curve shows the global mean spectrum. Dark and light shaded areas indicate $\pm\sigma$ from the mean spectrum and max/min range, respectively. Vertical dashed lines mark the intervals of the two studied peaks, K$\beta''$ and K$\beta_2$.}
    \label{fig:whole}
\end{figure}

\begin{figure*}[ht!]
\includegraphics[width=1.0\linewidth]{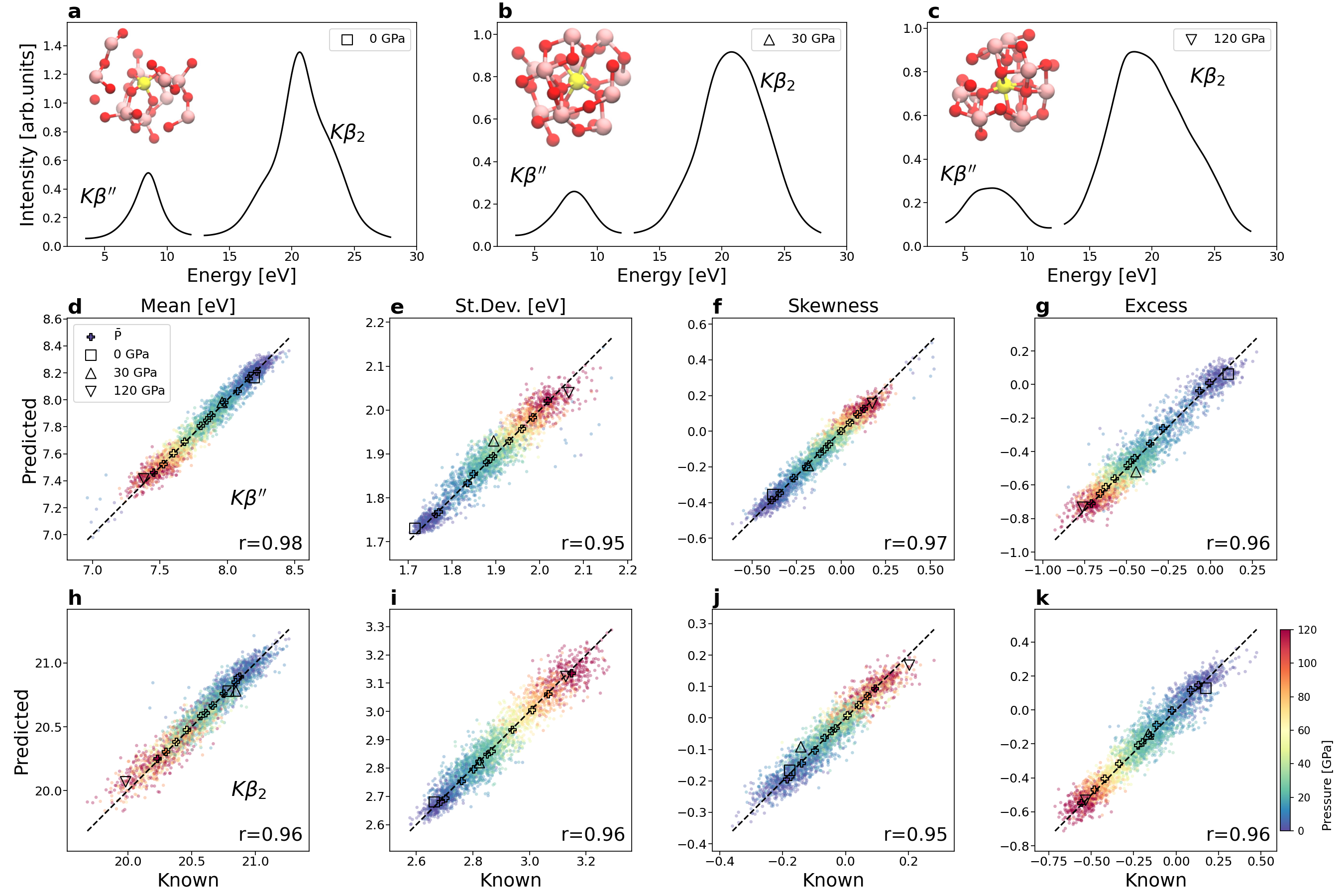}
\caption{(a--c)~Sample XES spectra at different pressures. For each spectrum the inset shows the corresponding 3D structure, where the active Ge site is yellow, other Ge sites are pink and O sites are red. The symbols are used to indicate the data point in the panels below. (d--k)~Results of the MLP training: predicted spectral moments of the evaluation data set for the K$\beta''$ (d--g) and K$\beta_2$ peak (h--k). The color of each point indicates the corresponding pressure for the structure, and colored crosses in every panel depict the mean values for each pressure subset (known: moments of known mean spectrum, predicted: mean of predicted moments). Positions of the sample spectra from panels (a--c) are marked with black markers. Number in every panel shows the Pearson's correlation coefficient $r$ between known and predicted data.}
\label{fig:results}
\end{figure*}
\section{Results}
The ensemble-averaged Ge K$\beta$ XES of GeO$_2$ shows a transition induced by pressure, as seen in Figure~\ref{fig:whole}. Even though a smooth progression of the spectrum as a function of pressure is observed, the underlying statistical variation in the condensed-phase XES is large \cite{Niskanen2016a,niskanen2019compatibility,Spiekermann2022}. This is manifested by gray shading in Figure \ref{fig:whole} that shows the minimum-maximum variation of intensity in the data set. We proceed with our analysis for the two lines with clear pressure dependence: the K$\beta''$ and the K$\beta_2$.
\par
Figure \ref{fig:results}a--c presents structures and spectra of three individual snapshots at pressures of 0, 30, and 120 GPa, respectively. Figures \ref{fig:results}d--k in turn show the prediction and training performance of the chosen MLP for these descriptors. In the figure, perfect match between known and predicted data lie on the diagonal dashed line. Furthermore, the positions of the three illustrated spectra of Figure \ref{fig:results}a--c, as well as the mean moment values for each pressure point against moment values of the known mean spectrum are indicated by crosses.
\par
The spectra and their statistical moments show a clear trend as a function of pressure. Moreover, the overall quality of the prediction performance yields Pearson correlation coefficients above 0.94. The pressure-induced progression in the spectra is transferred into spectral moments, for which the ML task proved to be easier than predicting spectra as vectors of channel-wise-listed intensity values (see Fig.~\ref{fig:momentsvsfull}). Analogously with simple intensity prediction, spectral moments of an ensemble-averaged spectrum can be estimated by the mean of predicted moments to a good accuracy (see crosses in Figure \ref{fig:results}d--k). However, this is an approximate finding instead of a mathematical equality.
\par
For the evaluation data set, some 77\% of generalized covered spectral variance ($R^2$ score) can be explained by only a single ECA component $\tilde{\mathbf{v}}_1$ (83\% with two components $\{\tilde{\mathbf{v}}_1,\tilde{\mathbf{v}}_2\}$). These components represent individually standardized elements of a Coulomb matrix unrolled to 153-dimensional vectors (for $\{\tilde{\mathbf{v}}_1,\tilde{\mathbf{v}}_2\}$ rolled back to the standardized Coulomb matrix differences, see Fig.~\ref{fig:ecavectors}) For PLSSVD, corresponding  spectral variance coverages were 73\% and 77\% for one and two components, respectively. The added contribution of the second component indicates a rapid drop of improvement in higher ranks.
\par
Before entering the inverse problem, it is instructive to analyse the decomposition of first rank {\it i.e.} along the path $\tilde{\mathbf{p}}^{(1)} = t_1\tilde{\mathbf{v}}_1$. Since the emulator provides the nonlinear response to the input vector, ECA is able to mimic the behavior of the moments more closely than PLSSVD, which is linear by definition (for the spectral moments along $t_1$ see Fig.~\ref{fig:moments_along_eca1}). For this reason, dimensionality reduction by ECA will be better adjusted to the spectral response; even with inaccuracies in prediction by the emulator, higher covered spectral variance is still obtained than from PLSSVD. For a majority of the atoms, both PLS and ECA trajectories follow the pressure-wise ensemble mean interatomic distances $R_{0i}$ from the active Ge site along the path (Fig.~\ref{fig:eca:proj}). 
However, for atoms Ge$_3$, Ge$_4$, O$_3$, O$_4$ and O$_7$ the ECA trajectories show a different behavior, which indicates that the role of these atoms in deciding the spectral outcome is low compared to other atoms.
\par
The interpretation of spectra would ideally lead to structures constructed from the spectroscopic information. However, already with a few number of degrees of freedom (here 153) this problem is tedious. In line with findings in Ref. \citenum{Niskanen2022neural}, training an emulator to directly predict the Coulomb matrix from the spectral moments was not successful with the model selection grid, data and descriptors used here (the mean Pearson correlation coefficient of 0.33 was obtained). Therefore we looked at approaches that would rely on spectrum prediction by an emulator, that has in general better performance. However, an emulator-based approach of iteratively fitting the parameters $\tilde{\mathbf{p}}$ to yield the 8 desired spectral moments proved also to be an unstable high-dimensional problem, that we were unable to solve. Instead, fitting a few ECA component scores for matching spectral moments is a much simpler task that could be solved.
\par
We searched for the coordinates $\mathbf{t}$ in the standardized dimension-reduced space by minimization of the least-squares error 
\begin{equation}\label{fiteq}
J(\mathbf{t})=\left|\left|\tilde{\mathbf{S}}_\mathrm{emu}\left({\sum}_{i=1}^k\ t_i\tilde{\mathbf{v}}_i\right)-\tilde{\mathbf{m}}\right|\right|^2
\end{equation}
for a data point $\tilde{\mathbf{p}}^{(k)}=\sum_{i=1}^kt_i\tilde{\mathbf{v}}_i$ with given standardized spectral moments $\tilde{\mathbf{m}}$. Here, $\tilde{\mathbf{S}}_\mathrm{emu}(\tilde{\mathbf{p}})$ is the standardized output of the moment emulator. We limit the study to two components $t_1$ and $t_2$. 
\par
\begin{figure*}[ht!]
\centering
\includegraphics[width=\linewidth]{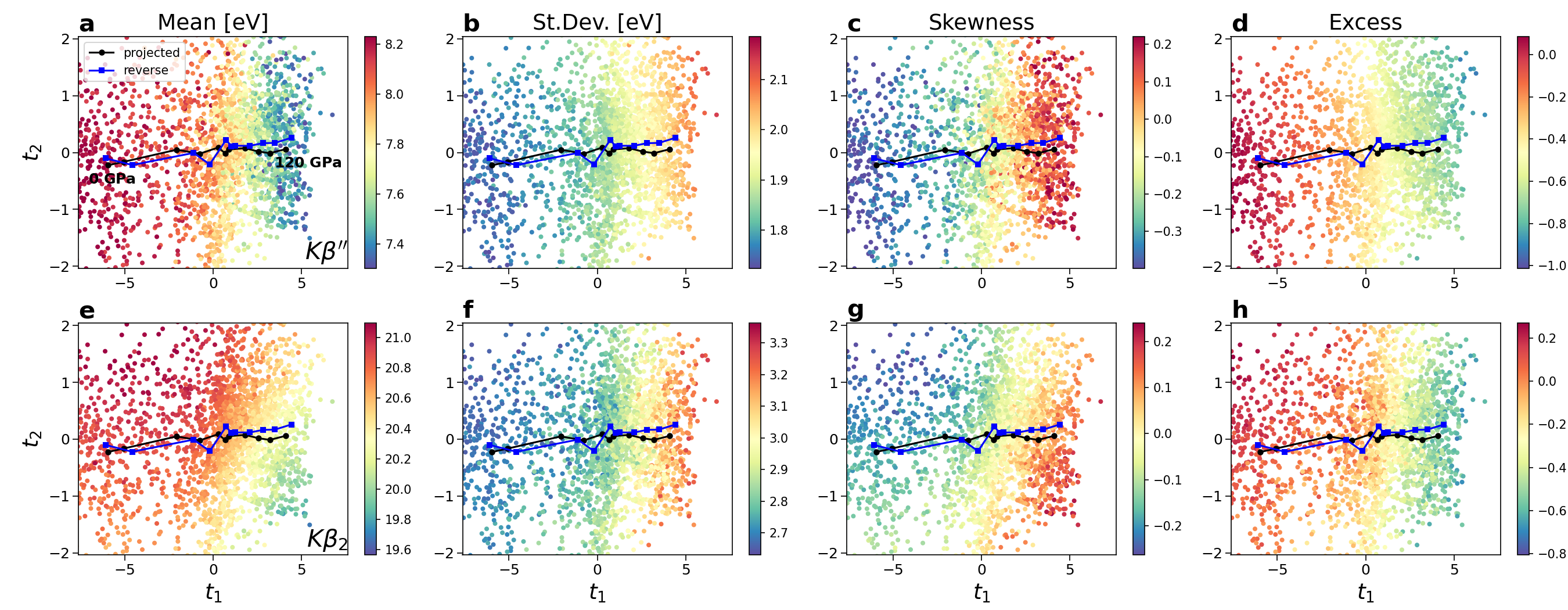}
\caption{Reconstructed $t_1$, $t_2$ coordinates for evaluation data. Individual data points are shown as colored markers, where color indicates the corresponding spectral moment. Black markers represent the projected mean coordinates for each pressure, and blue markers depict the reconstructed projections from the moments of ensemble-mean spectra}.
\label{fig:results:t1t2}
\end{figure*}
\par
Figure~\ref{fig:results:t1t2} shows coordinates $\mathbf{t}=(t_1,t_2)$ for each point from the evaluation data set from fitting of Equation~(\ref{fiteq}). Deduced coordinates for the set of moments of each mean-ensemble spectrum (blue line) are in a good agreement with projections of known mean points (black line) for a given pressure ensemble on the same subspace. It appears though, that this reconstruction of the scores $t_i$ misses the second component, possibly due to the fact that the component is already insignificant and the emulator is known to be imperfect. Knowledge of scores $t_i$ allows construction of an approximate Coulomb matrix $\tilde{\mathbf{p}}$ as a linear combination up to rank $k$. The absolute $\mathbf{p}$ is obtained after inverse standardization, as are $\mathbf{C}$ and $\mathbf{R}$. 
\par
\begin{figure}[!h]
\centering
\includegraphics[width=\linewidth]{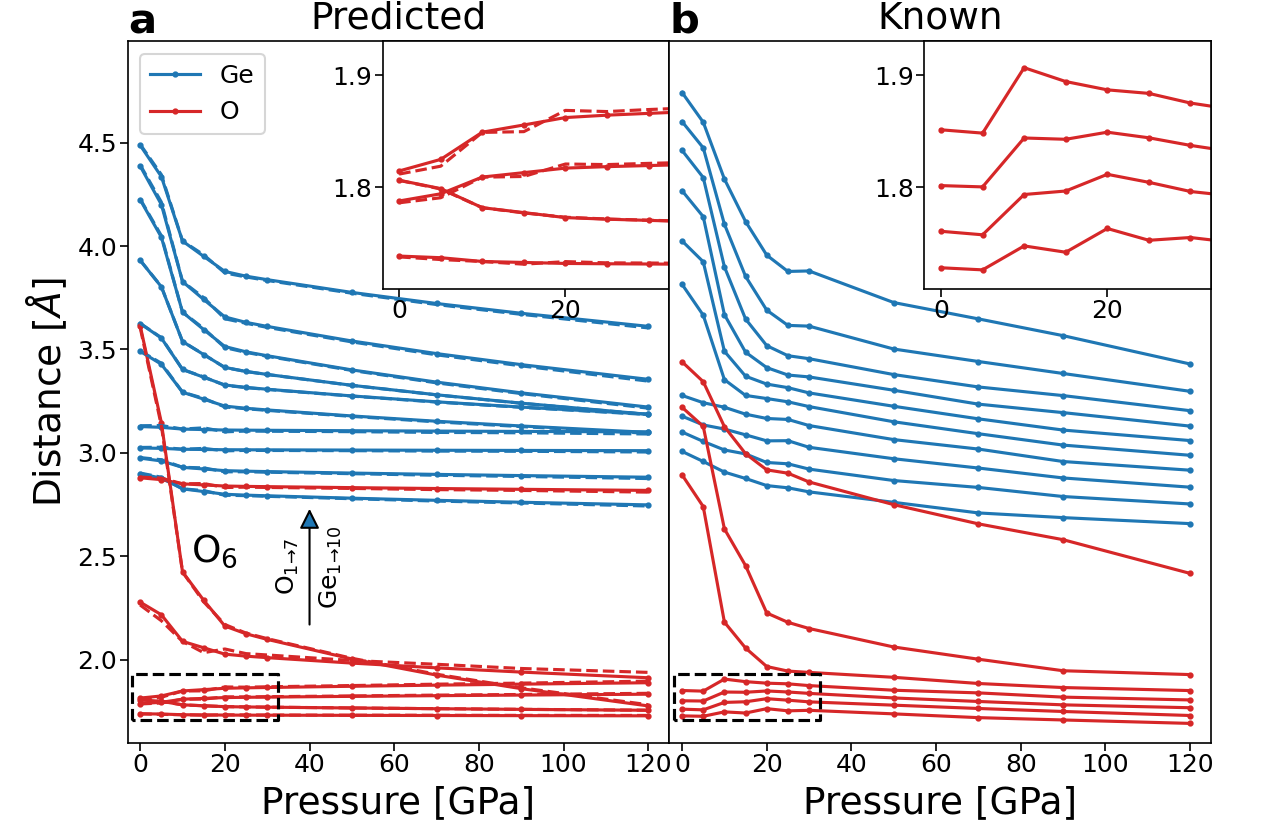}
\caption{Mean-structural-parameter-based distances from the central Ge atom. (a)~Reconstructed distances for the $t_1$ component. The inset shows distances for the 4 closest O atoms (dashed area). Dashed lines depict corresponding distances for sum of two projections ($t_1, t_2$). Arrow indicates the sequence of the atomic indices as mentioned in the text. (b)~Known mean distances evaluated from the atomic coordinates of the evaluation data set.}
\label{fig:t1projectionsbacktransformed}
\end{figure}

Even though the mean interatomic distances are not necessarily obtainable from mean Coulomb matrix elements, and even though this matrix is not necessarily obtainable from the spectral moments of the ensemble-mean spectrum (which closely match with the ensemble mean of the spectral moments), we find both to be the case. Figure \ref{fig:t1projectionsbacktransformed} depicts the reconstructed atomic distances from the spectral moments with one-dimensional and two-dimensional ECA space, indicating rapid convergence. Moreover, the reconstruction is at least qualitatively correct as seen from comparison with the known values for the evaluation data set, the most notable discrepancy being the 5th closest O atom at low pressures. This behavior can be understood in terms of reduced sensitivity of the spectra to these atomic distances; the first ECA component does not capture the drastic relative change in the parameter value (Fig.~\ref{fig:eca:proj}), and even the second component does not fix this shortcoming. Likewise, for the overall match on the data set,  the vector $\tilde{\mathbf{v}}_1$ results in underestimation of O$_6$ distance at large $t_1$ (high pressures), which leads to the line crossings in Figure~\ref{fig:t1projectionsbacktransformed}a.  However, the pressure-induced coordination change from 4-coordinated Ge to 6-coordinated Ge \cite{guthrie2004formation} is clearly discernible around the pressure of 10 GPa by the increase of the Ge--O separation for the first four oxygen atoms and the concomitant decrease of Ge--O distance for the fifth and sixth nearest oxygen neighbor. We note that while the first row of the constructed Coulomb matrix represent ensemble-averaged distances, the structure constructed from the mean Coulomb matrix is nonsensical.
\par

\section{Discussion}
With the limited data available it is essential to have structural and spectral descriptors that are linkable by rather simple MLPs. Consequently, the used descriptors dictate the analysis to follow. While the relative positions of atoms for a Coulomb matrix can be evaluated (see SI), there is dropout of more remote, potentially significant atoms. Furthermore, the observation that spectral moments are more suitable than tabulated intensities complicates spectral analysis as they may not be applicable in all cases, {\it e.g.} when clearly distinct and identifiable peaks are not formed for all data points.
\par
Instead of more direct approaches, approximate solution of the inverse problem by reconstruction of the first ECA components proved to be a feasible task to solve by optimization. It is natural to select these parameters so that they explain most spectral variance. As a result converging expansion of less and less relevant degrees of freedom are added and finally, irrelevant are identified and filtered out. Imperfection of emulator and incompleteness of the basis are likely reasons for the crossings of lines in Figure \ref{fig:t1projectionsbacktransformed}a.
\par
Structural analysis of the AIMD trajectory results in a complete analysis of structural changes across the data set. However, this information does not indicate what can be concluded based on the XES alone, as the sensitivity of core-level spectra to structural parameters may vary greatly\cite{Niskanen2022,Bergmann2020}. A parameter without an effect on a spectrum certainly cannot be expected to be reconstructed from it, and thus spectral insensitivity to a structural degree of freedom presents a danger of misinterpretation. The design of ECA means that a spectrally irrelevant structural degree of freedom obtains zero projection in the basis vector and is, in principle, omitted in subsequent analysis. Therefore, effects {\it shown} by ECA, and analysis based on it, can be considered to be inferred from a spectrum and its change. This reasoning is supported by Figure~\ref{fig:t1projectionsbacktransformed}, where the magnitudes of change from 0~GPa to 120~GPa in the known distance curves mostly exceed those of the predicted ones. For the end-to-end difference oxygens O$_3$ and O$_4$ with negligible ($<$0.05~{\AA}) total change exceed that of the known data. Depending on details of an analysis other -- rather minor -- violations to the tendency can be found in the data.
\par
For the 17 atoms and 11 pressures, the mean absolute deviation from the known ensemble-mean distances for 2-component decomposition was 0.091~\AA{} for ECA and notably 0.051~\AA{} for PLSSVD with which we also carried out the analysis (see figures \ref{fig:results:t1t2pls}--\ref{fig:t1plsprojectionsbacktransformed}). We interpret the better performance of PLSSVD to be due to more emphasis placed on structural variance in the method, whereas ECA focuses strictly on spectral significance. Thus PLS is allowed to know more from the simulated structural parameter space than the spectra alone would allow. However, the method undoubtedly benefited of the choice of descriptors by ML studies, making the data suitable for a linear model. In addition, imperfection of ECA results come from the imperfection of the emulator.
\par
Since the studied XES involves transitions of electrons from the occupied valence to localized deep core levels, the associated transition matrix elements become naturally limited to the immediate neighbourhood of the active atomic site. The occupied valence orbitals, in turn, can be expected to participate in chemical bonding, and thus to render these transitions sensitive to \textit{e.g.} coordination number of the active site. It is an interesting yet open question to which degree the findings presented here generalize in other systems, and specifically to those posed by XES of high-pressure science. When assuming no exceptionality for GeO$_2$ studied here, these spectra are potent of delivering far more structural information than it may at first seem.

\section{Conclusions}
For Ge K$\beta$ XES of GeO$_2$ at elevated pressures, Coulomb matrix and statistical moments of spectral peaks prove to be descriptors feasible to be linked by machine-learning applications with $\sim$10$^4$ simulated data points. We find the statistical moments of ensemble mean spectra to match closely with the ensemble mean of individually predicted moments. Dimensionality reduction by the ECA decomposition technique provides a means for a stable approximate solution of a spectroscopic inverse problem. We find the first row of a Coulomb matrix reconstructed from the spectral moments of the ensemble-mean spectrum to represent that obtained from ensemble-mean interatomic distances from the active site. Therefore, without a strict mathematical necessity, we find that these distances can be reconstructed to a good accuracy from the ensemble mean spectral statistical moments.
\par
Decomposition of structural sensitivity of spectra reduces the number of free parameters to be solved in the inversion problem, to only a few that have been chosen {\it a priori} for their spectral significance. The basis vectors of such decomposition span a subspace of degrees of freedom with most spectral response, and therefore reconstruction via this subspace will show structural effects with true inference from the change of spectra. This prevents spectrally irrelevant structural information, available in a simulation work, from affecting the analysis. Partial least squares fitting such as PLSSVD offers a usable and much lighter alternative where machine learning is not feasible, but the method is not as strict in spectrum-only inference.

\section*{Data availability}
Underlying data (including MD structures, corresponding XES spectra) and analysis scripts are available via request from the authors.

\section*{Author contributions}
A.V. data analysis, machine learning, writing the manuscript; C.J.S. simulations, data curation, writing the manuscript; J.N. research design, data curation and its assistive analysis, funding, writing the manuscript.

\section*{Conflicts of interest}
There are no conflicts to declare.

\section*{Acknowledgements}
Academy of Finland is acknowledged for funding via project 331234. The European Synchrotron Radiation Facility is thanked for providing  computing resources.

\bibliography{references}
\bibliographystyle{unsrt}

\clearpage
\onecolumngrid
\section*{Supporting Information}
\renewcommand\thefigure{S\arabic{figure}}    
\setcounter{figure}{0}
\section*{Optimal features configuration}
\begin{figure}[h!]
\centering
\includegraphics[width=\linewidth]{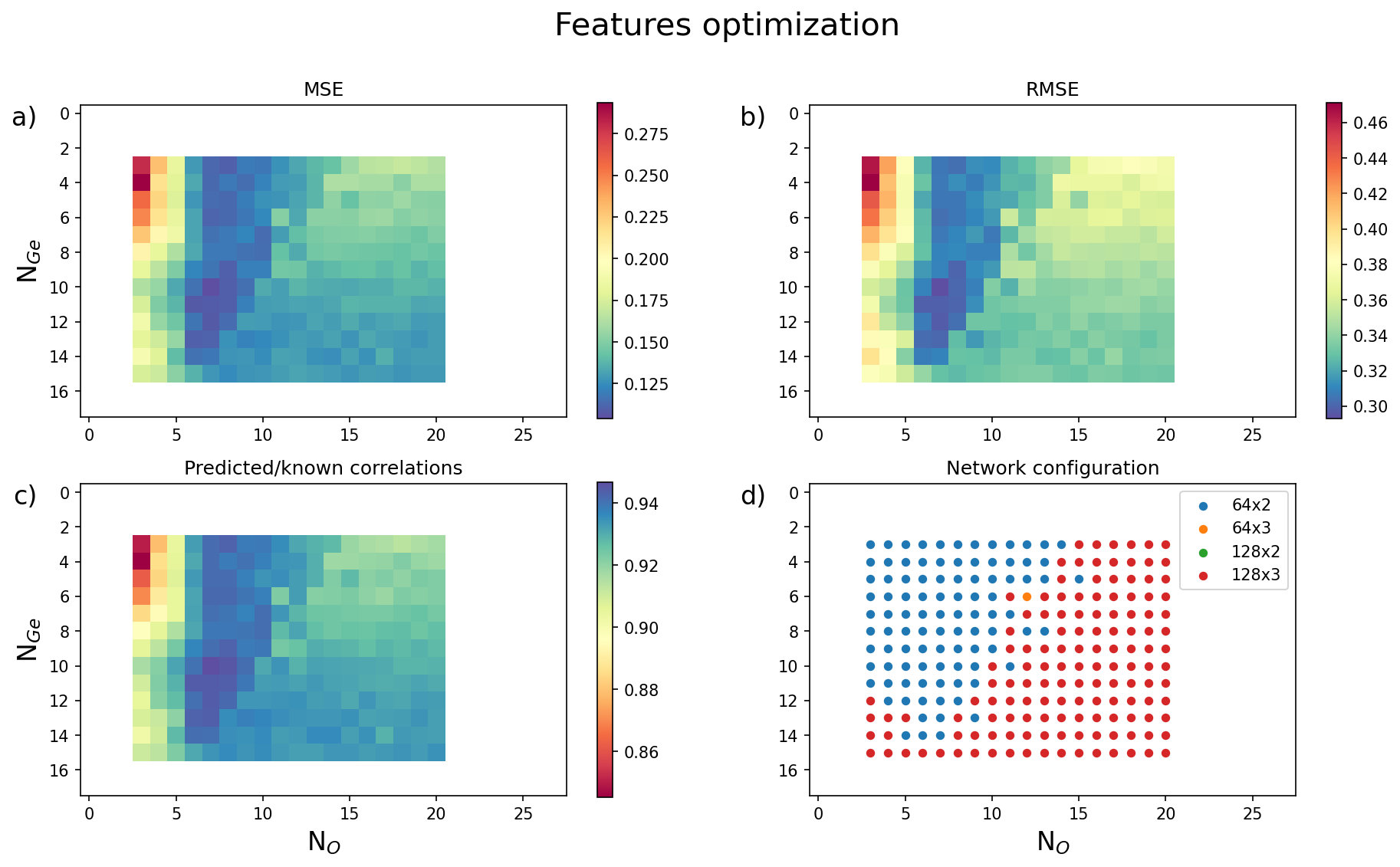}
\caption{Results of the features search. }
\label{fig:features}
\end{figure}

\begin{figure}[h!]
\centering
\includegraphics[width=\linewidth]{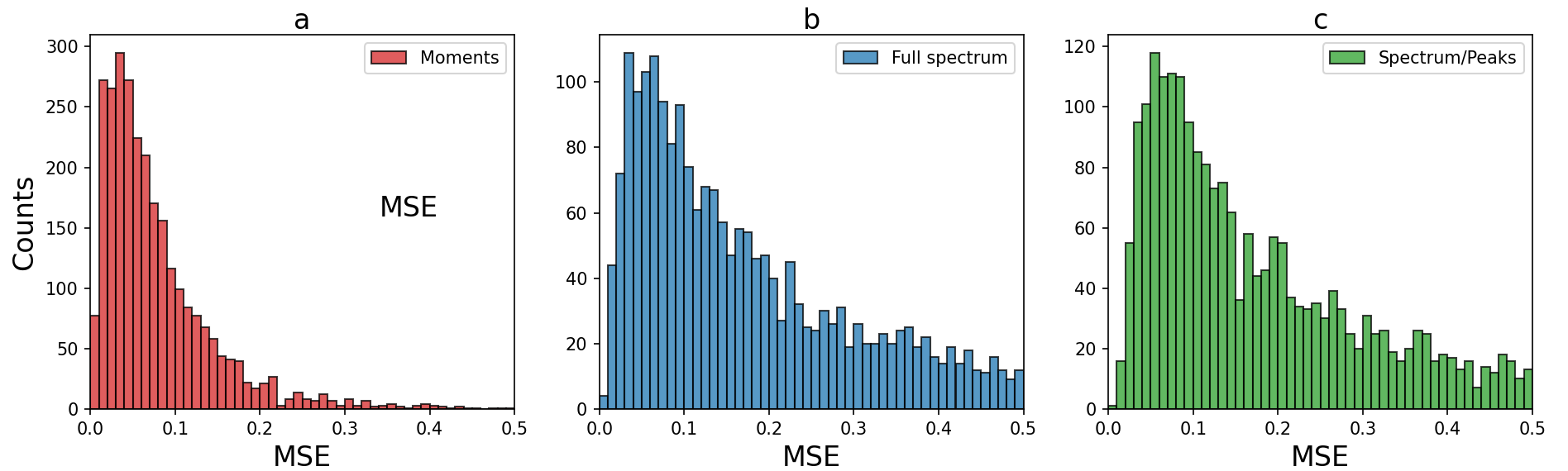}
\caption{Distribution of MSE values for a) 8 spectral moments prediction; b)~full spectrum prediction (500 pts); c)~spectrum within K$\beta''$--K$\beta_2$ peaks range prediction (3.5~eV--28~eV, 245 pts). }
\label{fig:momentsvsfull}
\end{figure}

\begin{figure}[h!]
\centering
\includegraphics[width=1.0\linewidth]{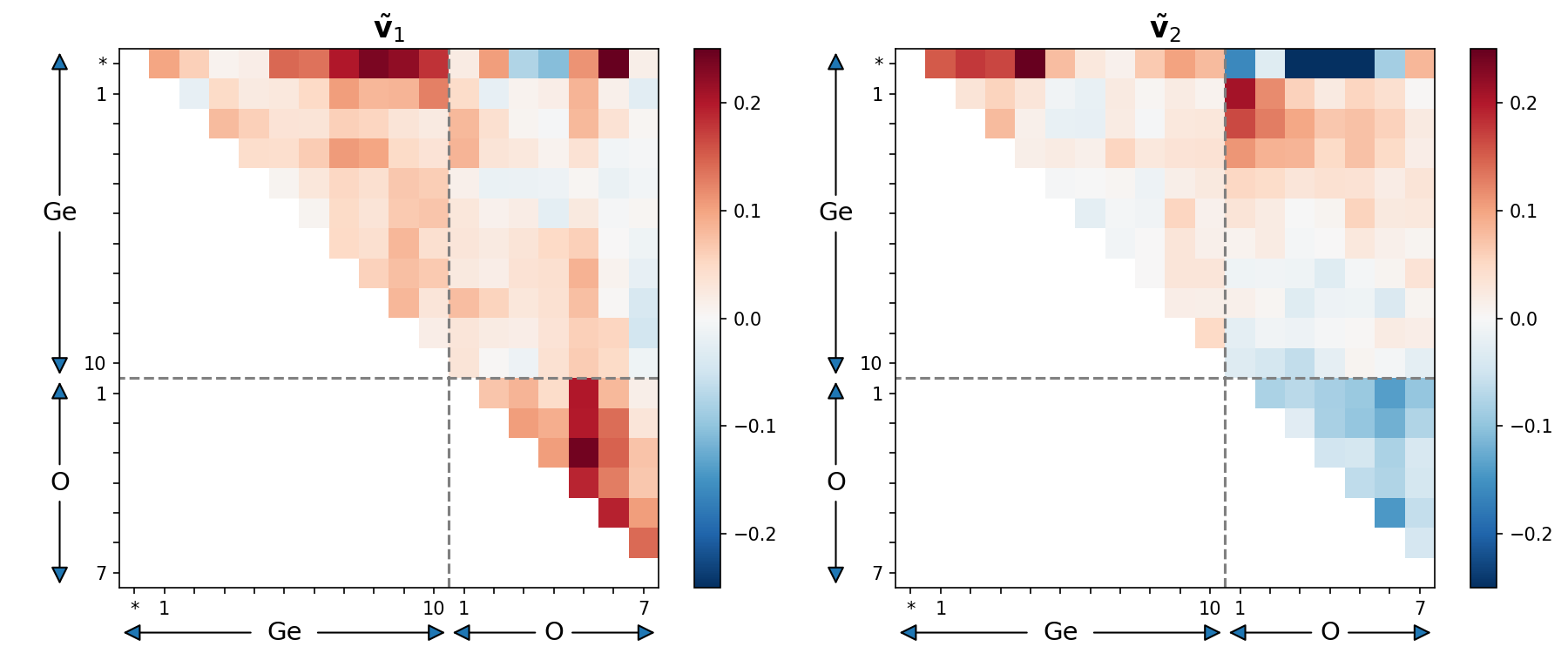}
\caption{First two ECA component vectors $\tilde{\mathbf{v}}_1$ and $\tilde{\mathbf{v}}_2$.}
\label{fig:ecavectors}
\end{figure}

To find the optimal number of Ge and O atoms for the best prediction of the spectral moments from the Coulomb matrices, we performed the grid search for each set containing of 3--15 Ge atoms and 3--20 O atoms, and for each MLP configuration with 2-3 hidden layers and 64 and 128 neurons in each layer, as presented in Figure~\ref{fig:features}. Panels a) and b) show mean MSE and mean RMSE for the predicted spectral moments. Panel c) shows the geometric mean of the pairwise correlations between known and predicted spectral moments. Panel d) depicts the configuration of the neural network with the best prediction for the given Ge+O set. 

For our system, the deduced optimal model contains 10 closest Ge atoms and 7 closest O atoms.

\newpage
\section*{Comparison of spectral moments prediction and full spectrum prediction}
We compared the prediction of the spectral moments and full spectrum for the same configuration of hidden layers in the neural network, the same structural features set configuration, and the same train/evaluation data splitting, using MSE of standardized spectral features as a metric. Additionally, we performed the comparison with a prediction of the spectrum within K$\beta''$--K$\beta_2$ peaks energy range. 

\newpage
\section*{Sample ECA components}
Every ECA component $\tilde{\mathbf{v}}$ is 153-dimensional vector in a space of normalized Coulomb matrices. Each component of ECA vector can be interpreted in terms of the changes in Coulomb matrix, which are attributed to the changes in the structure.

\newpage
\section*{Spectral moments along first component vectors}
\begin{figure*}[!h]
\centering
\includegraphics[width=\linewidth]{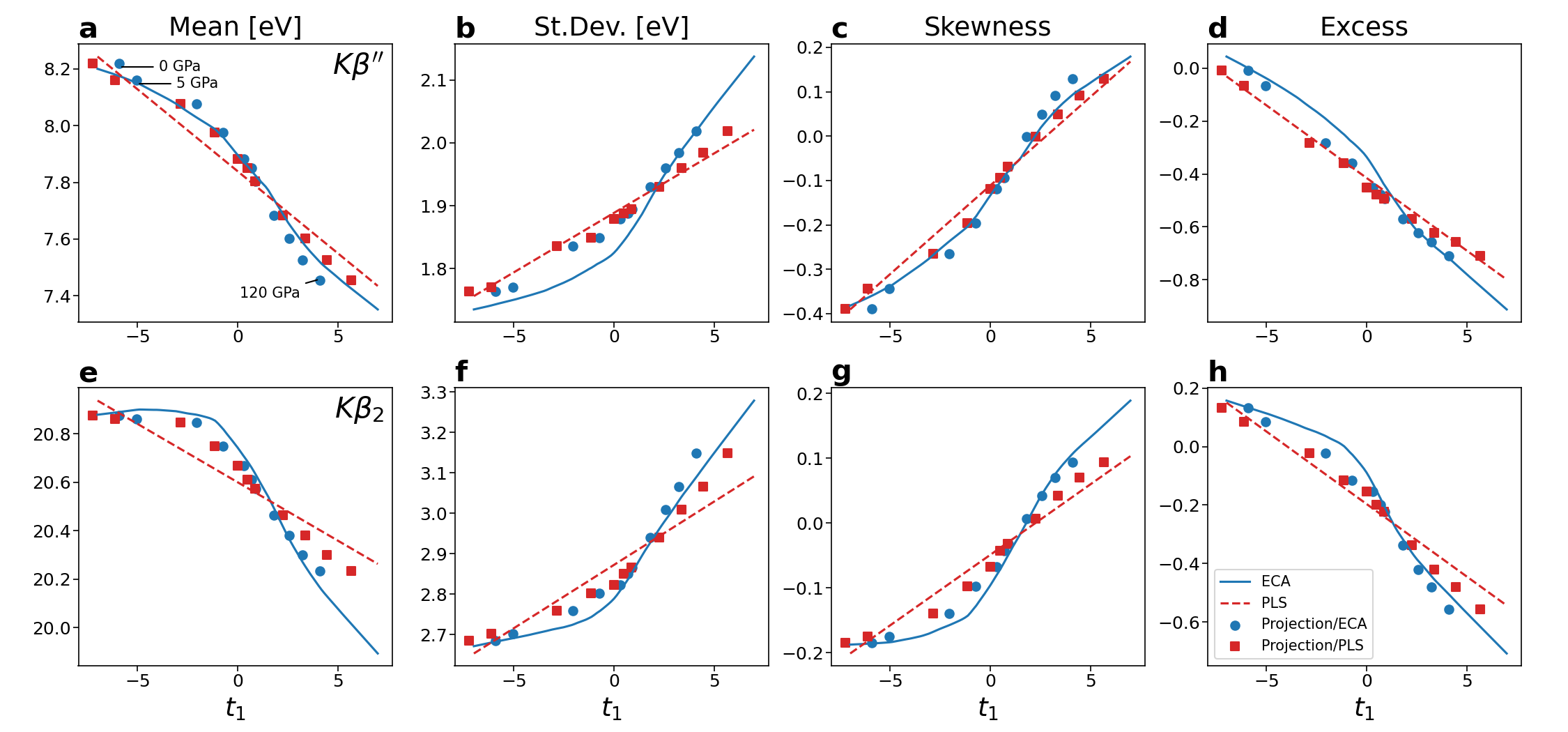}
\caption{Spectral moment predictions along the first decomposition components with the score $t_1$ range $[-7,7]$ to cover the known test data spread in the parameter space. Nonlinearity of the emulator makes ECA a nonlinear approach accounting for more spectral variance than PLSSVD. Blue and red markers represent the projections of every pressure mean on ECA and PLS first component vectors, respectively. The directions of the component vectors have been selected for the highest pressure to yield the highest score; in this case the pressure increases monotonously with the projection score. }
\label{fig:moments_along_eca1}
\end{figure*}

We evaluated the prediction of the spectral moments along the first ECA/PLS component vectors. For each $t_1$ value between $-7$ and $+7$, we calculated the normalized parameter space coordinates as $\tilde{\mathrm{p}} = t_1\tilde{\mathbf{v}}_1$ and $\tilde{\mathrm{p}} = t_1\tilde{\mathbf{v}}_1^\mathrm{(PLS)}$. The corresponding moments are deduced using model prediction and denormalization. These curves are shown as blue and red lines in Figure~\ref{fig:moments_along_eca1}. 
For each pressure, we also calculated the projection of the average parameter coordinate on ECA/PLS component vector, and average spectral moment. These ($t_1, m_i$) pairs are shown as colored markers.

\newpage
\section*{Projected distances from the central Ge atom to the nearest Ge and O atoms along first component vectors}
We analyzed how interatomic distances are changing along the component vectors for ECA and PLS approaches. Such analysis includes the following steps:

1) For each $t_1$ value between $-6$ and 6 we calculated the corresponding configuration in the normalized parameter space: $\tilde{p} = t_1\cdot \tilde{\mathbf{v}}_1$ (or $\tilde{\mathbf{v}}_1^\mathrm{(PLS)}$ for PLS).

2) Obtained 153-dimensional vector was inverse-standardized to deduce unrolled Coulomb matrix elements.

3) Coulomb matrix is converted into distance matrix, and its first row represents the distances from the central Ge atom (Ge$*$) to all other atoms used for Coulomb matrix calculation.

\begin{figure*}[!h]
\centering
\includegraphics[width=\linewidth]{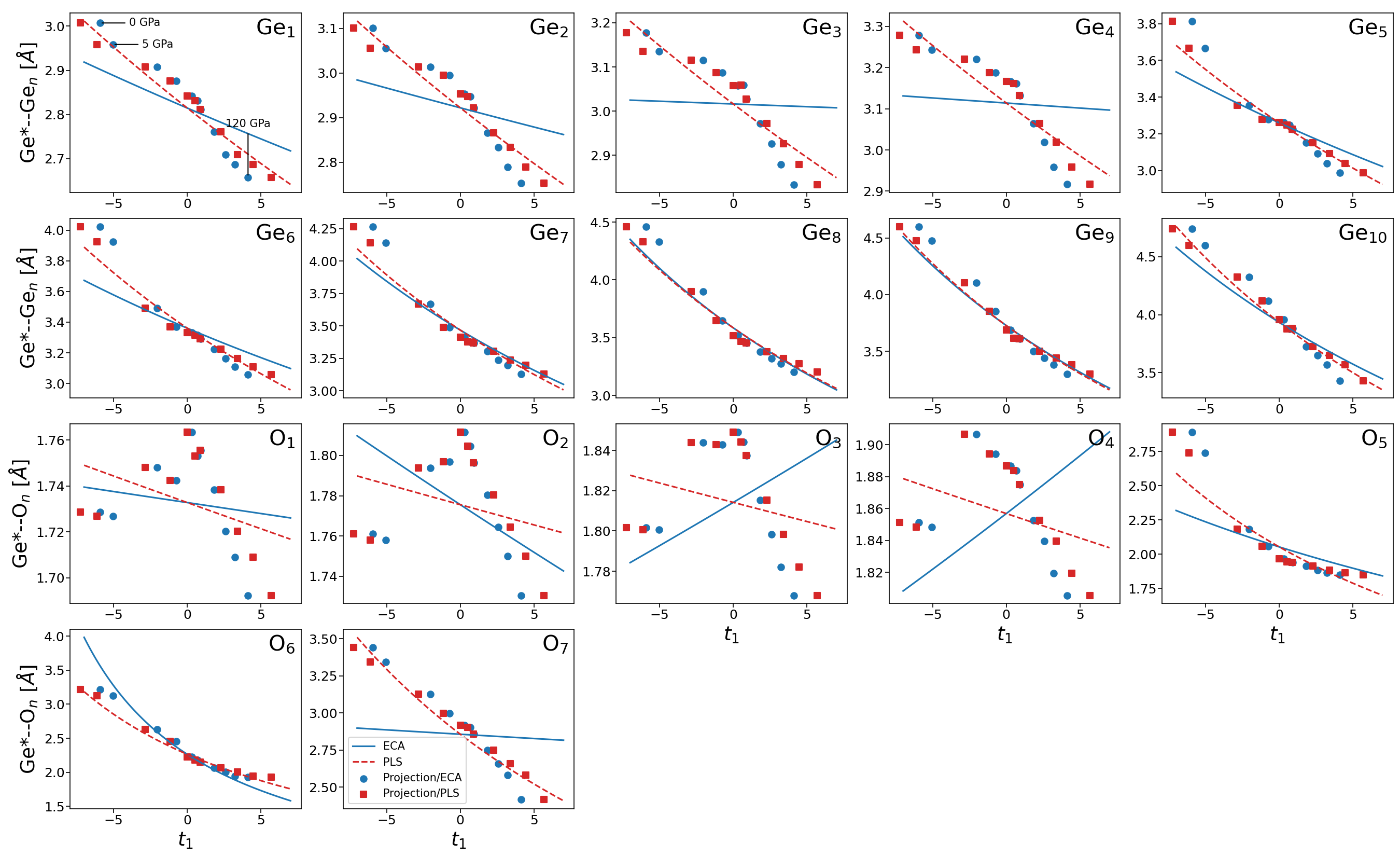}
\caption{Distances from the central Ge atom to the closest 10 Ge and 7 O atoms used for the calculation of the Coulomb matrix, deduced along the first ECA and first PLS component directions. Blue lines depict the  `predicted' positions of the atoms projected on the first EC component,  red dashed lines indicated the first PLS component. The projections of actual average distances for each pressure are shown as blue and red markers for ECA and PLS, respectively.}
\label{fig:eca:proj}
\end{figure*}

Obtained $d(t_1)$ dependencies are presented in Figure~\ref{fig:eca:proj}.

Also for each pressure, the projection of the mean scaled coulomb matrix on first component ECA/PLS vectors were calculated and presented as scatter plot in the same figure.

\newpage
\section*{Reconstruction of the spectral moments}
\subsection*{Reconstruction of the spectral moments from (t$_1$, t$_2$) projections in ECA space}
\begin{figure*}[!h]
\centering
\includegraphics[width=\linewidth]{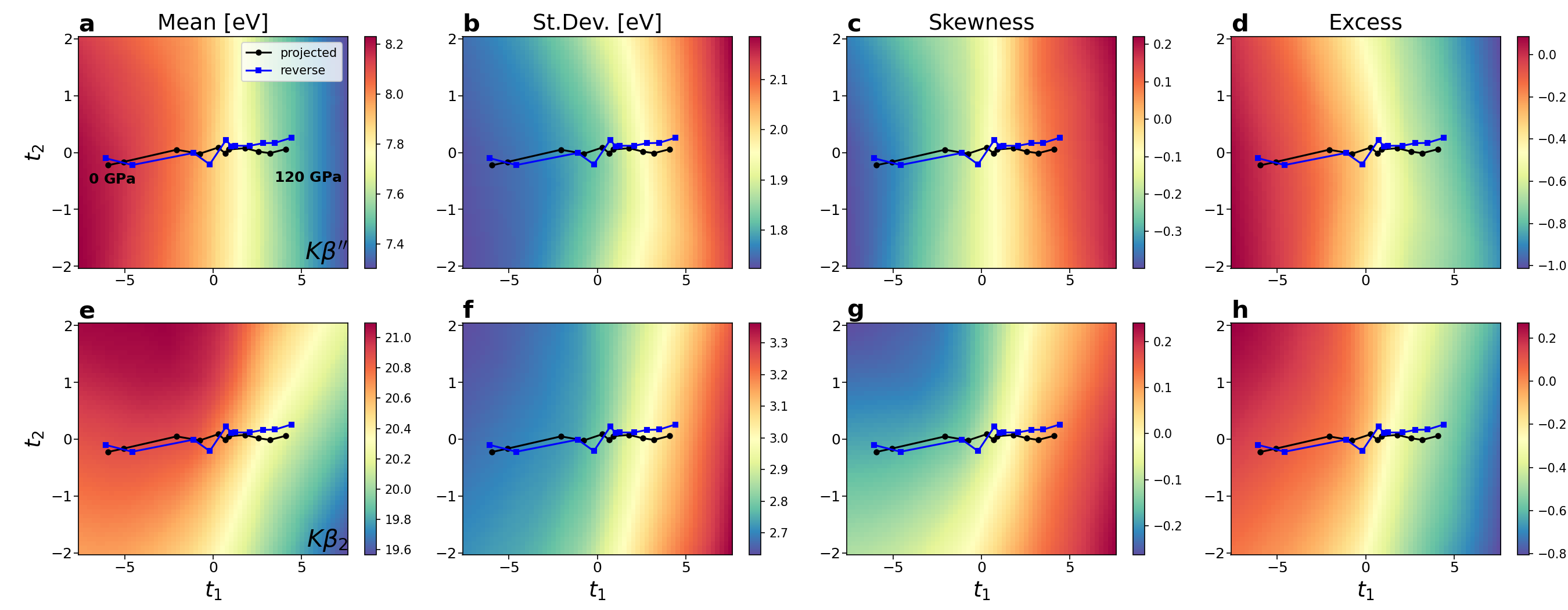}
\caption{Predicted spectral moments in 2-dimensional EC space. Each point on the plot represents a moment predicted by the model for the $\tilde{\mathbf{p}} = t_1\tilde{\mathbf{v}}_1+t_2\tilde{\mathbf{v}}_2$. Black markers represent the projected means parameters for each pressure, and blue markers depict the reconstructed projections from the mean moments.} 
\label{fig:results:t1t2projections}
\end{figure*}

Figure~\ref{fig:results:t1t2projections} shows the predicted spectral moments as a function in two-dimensional EC parameter space. Every point ($t_1, t_2$) on each of the 8 panels depicts the predicted spectral moment for the structure with $\tilde{\mathbf{p}} = t_1\tilde{\mathbf{v}}_1+t_2\tilde{\mathbf{v}}_2$.

\newpage
\subsection*{Reconstruction of the spectral moments from (t$_1$, t$_2$) PLS coordinates}
\begin{figure*}[ht!]
\centering
\includegraphics[width=\linewidth]{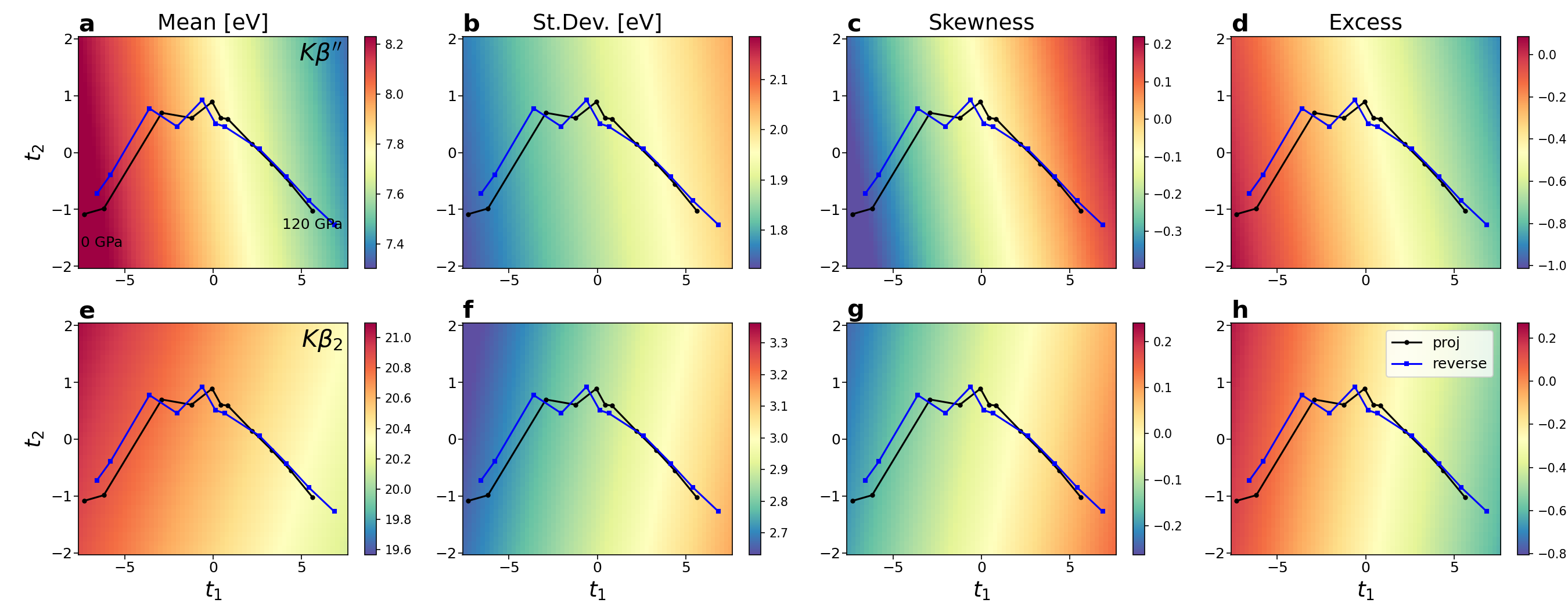}
\caption{Predicted spectral moments in 2-dimensional PLS space. Each point on the plot represents a moment predicted by the model for the $\tilde{\mathbf{p}} = t_1\tilde{\mathbf{v}}_1^\mathrm{(PLS)}+t_2\tilde{\mathbf{v}}_2^\mathrm{(PLS)}$. Black markers represent the projected means parameters for each pressure, and blue markers depict the inversely calculated projections from the mean moments.}
\label{fig:results:t1t2pls}
\end{figure*}

Figure~\ref{fig:results:t1t2pls} shows the predicted spectral moments as a function in two-dimensional PLS parameter space. Every point ($t_1, t_2$) on each of the 8 panels depicts the predicted spectral moment for the structure with $\tilde{\mathbf{p}} = t_1\tilde{\mathbf{v}}_1^\mathrm{(PLS)}+t_2\tilde{\mathbf{v}}_2^\mathrm{(PLS)}$.

\begin{figure*}[!h]
\centering
\includegraphics[width=\linewidth]{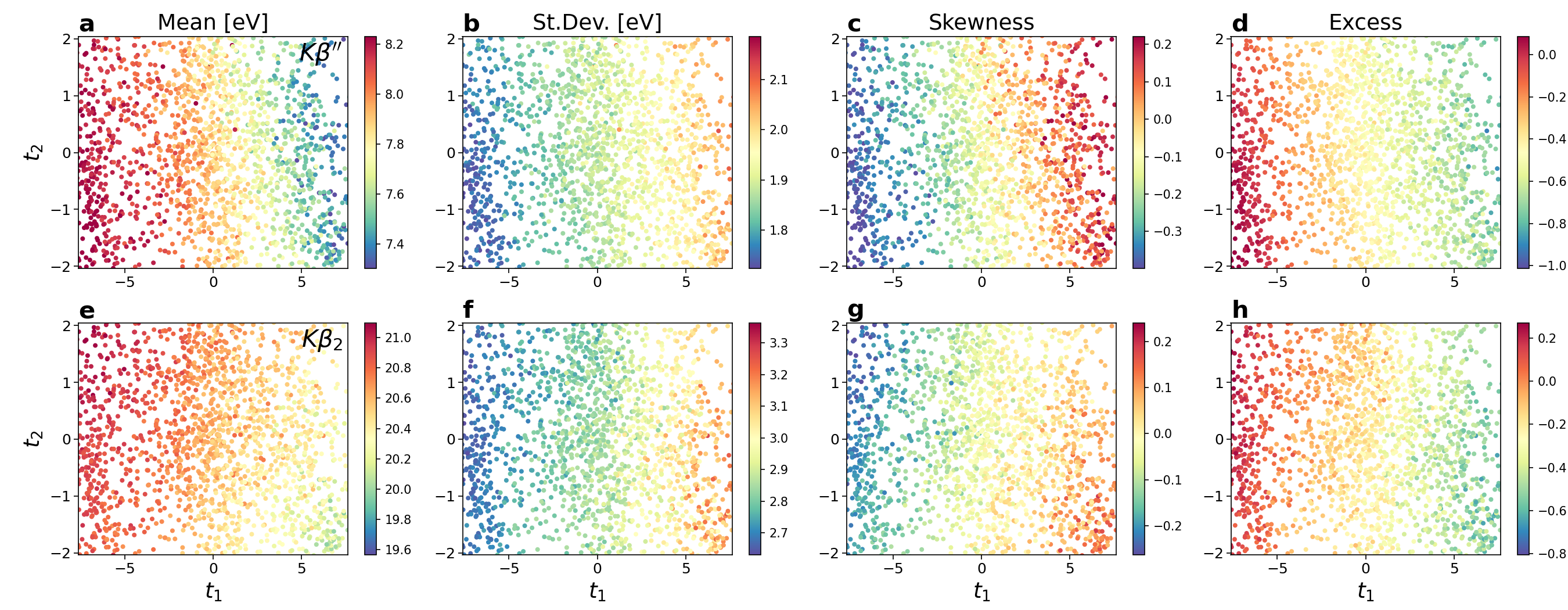}
\caption{Reconstructed $t_1$, $t_2$ coordinates for all evaluation data points using PLS approach. Color scales are identical for each corresponding panel in the Figure~\ref{fig:results:t1t2pls}.}
\label{fig:results:t1t2_plsprojections}
\end{figure*}

\begin{figure}[!h]
\centering
\includegraphics[width=0.8\linewidth]{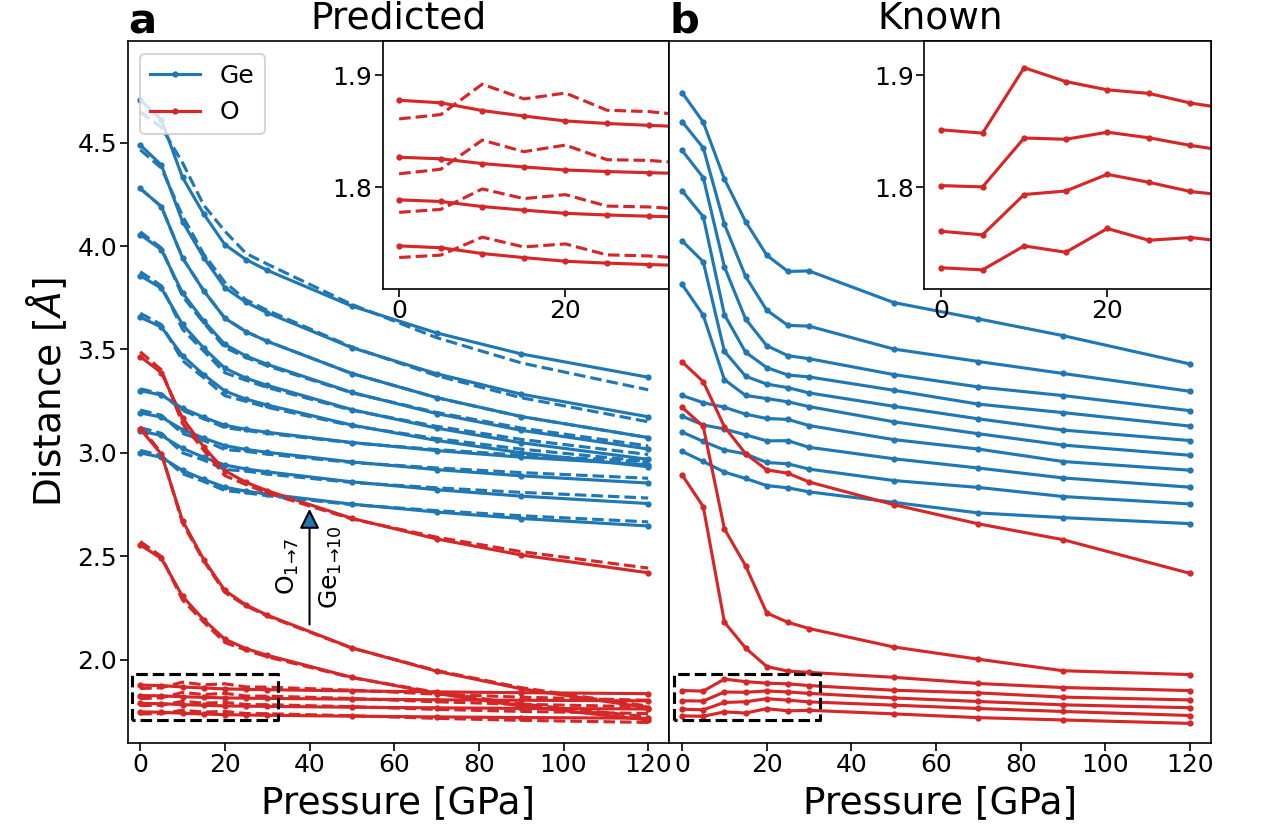}
\caption{(a)~Reconstructed mean-structural-parameter-based distances from the central atom for $t_1^\mathrm{(PLS)}$ component. The inset shows distances for the 4 closest O atoms (dashed area). Dashed lines depict corresponding distances for sum of two projections ($t_1, t_2$). (b)~Known mean-ensemble distances. }
\label{fig:t1plsprojectionsbacktransformed}
\end{figure}

Figure~\ref{fig:results:t1t2_plsprojections} shows the reconstructed ($t_1, t_2$) coordinates from spectral moments using PLS approach, for each point from evaluation data set. 

\newpage
\section*{Reconstruction of the interatomic distances as a function of pressure using PLS projections} 
Analysis of interatomic distances for the PLSSVD was done using the same approach as for Figure~5 im the main text. 

\newpage
\section*{Predicted atomic distances from ECA projections}
To evaluate the validity of predicted interatomic distances, we compared predicted curves with known values. 
\begin{figure}[h!]
    \centering
    \includegraphics[width=0.9\linewidth]{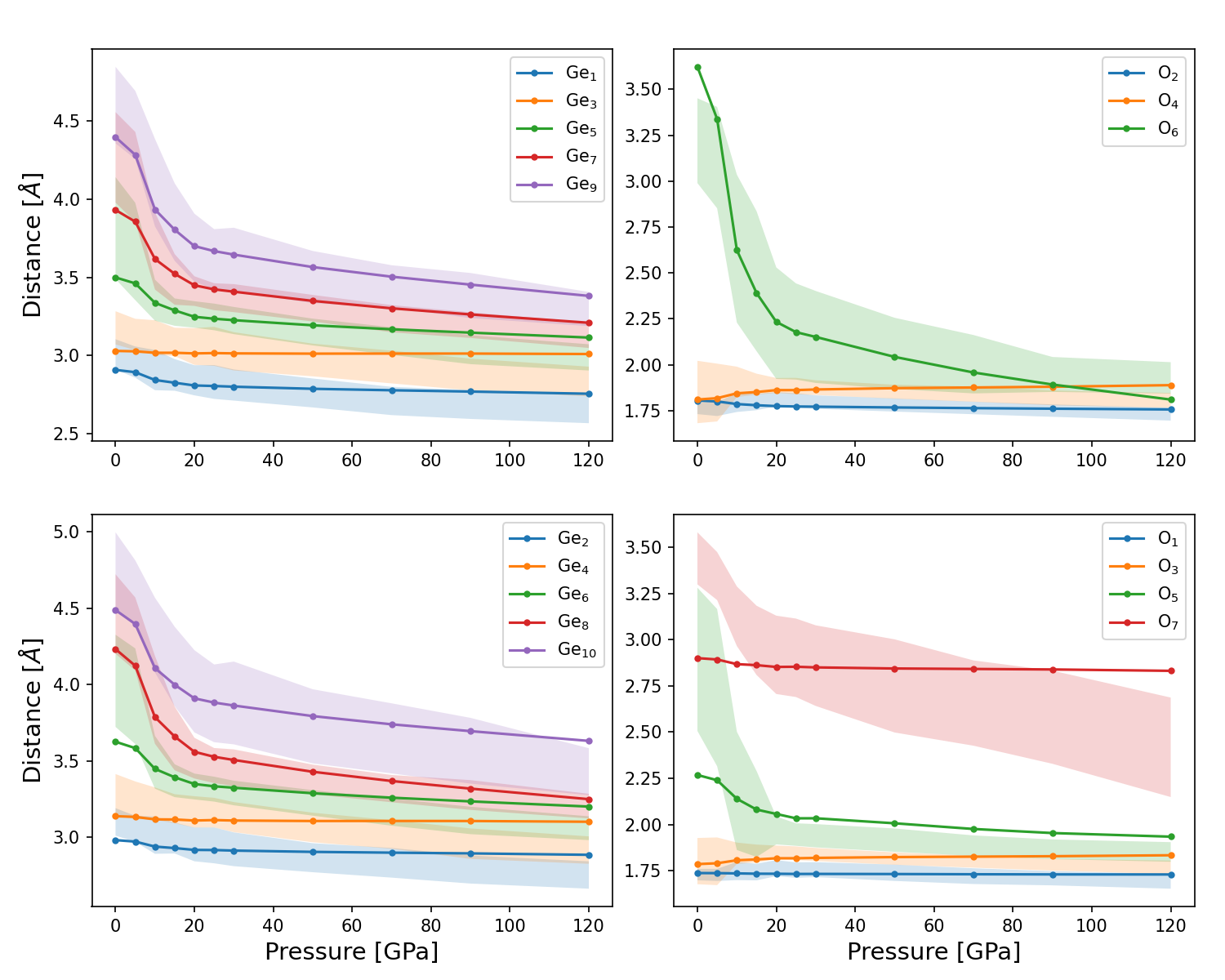}
    \caption{Distances from the central Ge atom to the neighboring 10 germanium atoms and 7 oxygen atoms. Shaded areas: mean$\pm$std of the known distances for each pressure. Lines: projected distance for each mean spectral moment per pressure.}
    \label{fig:ddproj}
\end{figure}

\FloatBarrier
\newpage

\section*{Reconstruction of structure from the Coulomb matrix}
We start indexing of atoms from 0 and denote the cartesian coordinates of atomic site $n$ by $(x_n,y_n,z_n)$. The original snapshot can be recovered from the distance matrix $\mathbf{R}$ (and from the Coulomb matrix $\mathbf{C})$ by using the following algorithm
\begin{enumerate}
\item Place the active site 0 at origin
\item Place site 1 at $(R_{01},0,0)$
\item Place site 2 at $(R_{02}\cos\gamma,R_{02}\sin\gamma,0)$ (in xy-plane in 
positive direction) where
\begin{eqnarray}
\gamma & = & \arccos{\frac{R_{01}^2+R_{02}^2-R_{12}^2}{2R_{01}R_{02}}}
\end{eqnarray}
\item Place site 3 in positive $z$ (negative $z$ for the alternate handedness) so that 
$R_{03},R_{13},R_{23}$ are fulfilled.
This corresponds to solving
\begin{equation}
-2\begin{bmatrix}
x_1 & y_1 & z_1 \\
x_2 & y_2 & z_2 
\end{bmatrix}
\begin{bmatrix}
x_3\\y_3\\z_3
\end{bmatrix}
=
\begin{bmatrix}
(R_{13}^2 - R_{03}^2 - R_{01}^2) \\ 
(R_{23}^2 - R_{03}^2 - R_{02}^2) \\
\end{bmatrix}
\end{equation}
by the Moore-Penrose pseudoinverse to obtain $(x_3,y_3,0)$. The component $z_3$ is then obtained as
\begin{equation}
    z_3 = \sqrt{R_{03}^2-x_3^2-y_3^2},
\end{equation}
where the requirement for positive $z_3$ (negative $z_3$ for the alternate handedness) has been used to obtain the unique solution.

\item Place each remaining site $n$ so that 
$R_{0n},R_{1n},R_{2n},R_{3n}$ are fulfilled. This corresponds to solving
\begin{eqnarray}
-2\begin{bmatrix}
x_1 & y_1 & z_1 \\
x_2 & y_2 & z_2 \\
x_3 & y_3 & z_3
\end{bmatrix}
\begin{bmatrix}
x_n\\y_n\\z_n
\end{bmatrix}
=
\begin{bmatrix}
(R_{1n}^2 - R_{0n}^2 - R_{01}^2) \\ 
(R_{2n}^2 - R_{0n}^2 - R_{02}^2) \\
(R_{3n}^2 - R_{0n}^2 - R_{03}^2)
\end{bmatrix}
\end{eqnarray}
with respect to $(x_n,y_n,z_n)$ which are the coordinates of atom $n$.

\end{enumerate}
Steps 1-4 fix the orientation and handedness of the coordinate system. An analysis of all structural data confirms, that apart from this handedness, each constructed structure matches with the original one after a suitable rotation (maximal location deviation of the order 10$^{-8}$~{\AA}). This also means that only the first 4 rows of the upper triangle of the symmetric Coulomb matrix describe the geometry completely apart from the aforementioned handedness-based symmetry. A code of the reconstruction algorithm is given with the data and other scripts.

\end{document}